\documentclass[aps,showpacs,twocolumn,preprintnumbers,bibnotes]{revtex4-1}
\usepackage{ifpdf}
\usepackage{graphicx}
\usepackage{color}
\usepackage{dcolumn}
\usepackage{bm}
\usepackage{mathrsfs}
\usepackage{amsfonts}
\usepackage{amssymb}
\usepackage{amsmath}
\usepackage{dsfont}
\usepackage{longtable}
\usepackage{fixmath}
\usepackage{upgreek}
\usepackage{latexsym}
\usepackage{bbding}
\usepackage{bold-extra}
\usepackage[T1]{fontenc}
\usepackage{ae,aecompl}
\usepackage{url}
\usepackage[dvipsnames*,svgnames]{xcolor}
\usepackage[pdfauthor={Behnam Farid},%
            pdftitle={Comment on ``Breaking the theoretical scaling limit for predicting quasi-particle energies: The
stochastic GW approach'', by Daniel Neuhauser et al., arXiv:1402.5035v1},%
            plainpages=false,pdfpagelabels,pagebackref,%
            breaklinks=true,%
            hyperfootnotes=true,%
            bookmarks=true]{hyperref}
\hypersetup{colorlinks=true,%
            citecolor=blue,%
            filecolor=black,%
            linkcolor=blue,%
            urlcolor=blue,%
            pdftex,%
            pdfstartview={FitH}}
\usepackage[all]{hypcap}
\makeatletter
\renewcommand{\@fnsymbol}[1]{\ensuremath{\ifcase#1\or *\or \S \else\@ctrerr\fi}}
\makeatother
\def\t#1{\tilde{#1}}
\def\wh#1{\widehat{#1}}
\def\h#1{\hat{#1}}
\DeclareMathOperator{\re}{Re}
\DeclareMathOperator{\im}{Im}
\DeclareMathOperator{\e}{e}
\DeclareMathOperator{\rd}{d\!}
\DeclareMathOperator{\sgn}{sgn}
\begin{document}
\ifx\href\undefined\else\hypersetup{linktocpage=true}\fi
\bibliographystyle{apsrev}

\title{Comment on ``Breaking the theoretical scaling limit for predicting quasi-particle energies: The
stochastic $GW$ approach'', by Daniel Neuhauser \emph{et al.}, \href{http://arxiv.org/abs/1402.5035}{arXiv:1402.5035v1}}
\author{Behnam Farid}
\protect\email{behnam.farid@btinternet.com}

\date{\today}

\begin{abstract}
We show that the recently-introduced formalism by Neuhauser \emph{et al.} for the calculation of the quasi-particle energies of electronic systems within the framework of the $GW$ approximation of the self-energy operator, named the `stochastic $GW$ approach' and empirically shown to have a linear-scaling arithmetic complexity for increasing number of electrons, suffers from two fundamental shortcomings that \textsl{cannot} be overcome while maintaining the present empirical linear-scaling property of the approach.
\end{abstract}


\maketitle
In a recent publication, Neuhauser \emph{et al.} \cite{NGAKRB14a,NGAKRB14b} have presented a formalism for the calculation of the quasi-particle energies of interacting electronic systems within the framework of the $GW$ approximation \cite{LH65,HL69} of the self-energy operator. Numerical calculations by the authors, of the highest occupied and the lowest unoccupied quasi-particles energy levels of a series of hydrogen-passivated silicon nanocrystals on finite grids, have shown that the required CPU time increases almost linearly with increasing number of electrons. This approach relying on some stochastic computational methods, Neuhauser \emph{et al.}  \cite{NGAKRB14a} have named it the `stochastic $GW$ [s$GW$] approach'. In this Comment we shall \textsl{not} discuss the stochastic elements of the s$GW$ formalism and restrict our considerations to what we show to be two of its \textsl{fundamental} shortcomings. Here we suffice to mention that the stochastic closure relation $\h{1} = \langle \vert \phi\rangle\langle\phi\vert\rangle_{\phi}$ as utilized in Ref.\,\cite{NGAKRB14a}, is equivalent to the orthonormality relationship that underlies the stochastic method of matrix inversion, such as investigated by Dong and Liu \cite{DL94} (see Eq.\,(1) herein) and adopted by Krajewski and Parrinello \cite{KP05}, which relies on $Z_2$-distributed random noise vectors. Our investigations, to be published, reveal that this method is extremely inefficient in the electronic-structure calculations regarding realistic periodic systems, specifically when these calculations are to be performed to self-consistency \cite{BF14}.

Calculating the self-energy operator $\h{\Sigma}_{\sigma}(\varepsilon)$ for particles with spin index $\sigma$ and at energy $\varepsilon \equiv \hbar\omega$ to first order in the screened interaction operator $\h{W}$ \cite{JH57} in terms of the single-particle Green operators $\{\h{G}_{\sigma}^{\textsc{ks}}(\varepsilon)\| \sigma\}$ corresponding to the $N$-particle non-interacting ground state (GS) of the many-body Kohn-Sham Hamiltonian $\wh{H}_{\textsc{ks}}$ \cite{KS65}, one obtains the non-self-consistent $GW$ self-energy operator $\h{\Sigma}_{\sigma}^{\textsc{gw}}(\varepsilon)$ \cite{LH65,HL69} minus the exchange-correlation-potential contribution $\hbar^{-1} \h{v}_{\sigma}^{\textrm{xc}}[\{n_{\sigma'}\}]$; the Hartree-potential contribution $\hbar^{-1} \h{v}^{\textsc{h}}[n]$, which is a functional of the total number density $n(\bm{r}) = \sum_{\sigma} n_{\sigma}(\bm{r})$, is exactly cancelled through this potential being included in the single-particle Kohn-Sham Hamiltonian $\h{h}_{\sigma}[\{n_{\sigma'}\}]$, Eq.\,(\ref{e3}), describing $\wh{H}_{\textsc{ks}}$ \cite[\S7.6, p.\,182]{BF99a}.

For clarity, with the exception of $\wh{H}_{\textsc{ks}}$, which denotes a many-body operator, in this Comment \textsl{all} symbols with a caret placed above them denote single-particle operators acting in the single-particle Hilbert space of the system under consideration.

With $\h{W}(\varepsilon) \equiv \h{v} + \h{W}'(\varepsilon)$, where $\h{v}$ is the bare two-body interaction potential operator, and $\h{W}'(\varepsilon)$ the potential operator arising from the dynamical screening of electrons \cite{JH57}, $\h{\Sigma}_{\sigma}^{\textsc{gw}}(\varepsilon)$ can be expressed as
\begin{equation}\label{e1}
\h{\Sigma}_{\sigma}^{\textsc{gw}}(\varepsilon) \equiv \h{\Sigma}_{\sigma}^{\textsc{g}v} + \h{\Sigma}_{\sigma}^{\textsc{gw}'\hspace{-2.0pt}}(\varepsilon),
\end{equation}
where $\h{\Sigma}_{\sigma}^{\textsc{g}v}$ is the static exchange self-energy operator, encountered in the Hartree-Fock theory (denoted by $\h{\Sigma}^{\textsc{x}}$ in Ref.\,\cite{NGAKRB14a}), for which one has
\begin{equation}\label{e2}
\langle\bm{r}\vert\h{\Sigma}_{\sigma}^{\textsc{g}v}\vert\bm{r}'\rangle = -\frac{1}{\hbar} v(\bm{r}-\bm{r}') \varrho_{\sigma}(\bm{r},\bm{r}'),
\end{equation}
where $v(\bm{r}-\bm{r}') \equiv \langle\bm{r}\vert\h{v}\vert\bm{r}'\rangle$, and $\varrho_{\sigma}(\bm{r},\bm{r}') \equiv \langle\bm{r}\vert\h{\varrho}_{\sigma}\vert\bm{r}'\rangle$ is the single-particle density matrix. Since here the self-energy operator is determined in terms of $\{\h{G}_{\sigma}^{\textsc{ks}}(\varepsilon)\| \sigma\}$, the function $\varrho_{\sigma}(\bm{r},\bm{r}')$ in Eq.\,(\ref{e2}) is the Kohn-Sham single-particle density matrix, which is idempotent.

With $\h{h}_{\sigma}[\{n_{\sigma'}\}]$ denoting the single-particle Kohn-Sham Hamiltonian corresponding to electrons with spin index $\sigma$, $\sigma \in \{\uparrow,\downarrow\}$, one has \cite{KS65,vBH72}
\begin{equation}\label{e3}
\h{h}_{\sigma}[\{n_{\sigma'}\}] = \frac{\h{p}^2}{2\mathsf{m}} + \h{u} + \h{v}^{\textsc{h}}[n] + \h{v}_{\sigma}^{\textrm{xc}}[\{n_{\sigma'}\}],
\end{equation}
where $\h{\bm{p}}$ ($\h{p}^2 \equiv \h{\bm{p}}\cdot \h{\bm{p}}$) is the single-particle momentum operator, $\mathsf{m}$ the electron mass, and $\h{u}$ the operator for the external potential. Suppressing $[n]$ and $[\{n_{\sigma'}\}]$, for the inverse of $\h{G}_{\sigma}^{\textsc{ks}}(\varepsilon)$ one has
\begin{equation}\label{e4}
\hbar\hspace{0.4pt}\h{G}_{\sigma}^{\textsc{ks}\hspace{0.4pt}-1}(\varepsilon) = \varepsilon \h{1} - \h{h}_{\sigma},
\end{equation}
so that, following the Dyson equation \cite{FW03}, for the inverse of the interacting single-particle Green operator $\h{G}_{\sigma}(\varepsilon)$ within the framework of the non-self-consistent $GW$ approximation one obtains
\begin{equation}\label{e5}
\hbar\hspace{0.4pt}\h{G}_{\sigma}^{-1}(\varepsilon) = \varepsilon \h{1} - \big(\h{h}_{\sigma} - \h{v}_{\sigma}^{\textrm{xc}} +  \hbar\hspace{0.4pt}\h{\Sigma}_{\sigma}^{\textsc{g}v} + \hbar\hspace{0.4pt}\h{\Sigma}_{\sigma}^{\textsc{gw}'\hspace{-2.0pt}}(\varepsilon)\big).
\end{equation}
Above $\h{1}$ denotes the identity operator in the single-particle Hilbert space of the problem at hand.

The single-particle excitation energies of the interacting system are the energies $\varepsilon$ for which $\h{G}_{\sigma}(\varepsilon)$ is unbounded (more about this later). To stay close to the treatment in Ref.\,\cite{NGAKRB14a}, we consider the matrix representations of the above operators with respect to the normalized single-particle eigenstates of the Kohn-Sham Hamiltonian $\h{h}_{\sigma}[\{n_{\sigma'}\}]$, Eq.\,(\ref{e3}), that is $\{\vert\psi_{\sigma;i}\rangle \| i\}$, for which one has
\begin{equation}\label{e6}
\h{h}_{\sigma}[\{n_{\sigma'}\}] \vert\psi_{\sigma;i}\rangle = \varepsilon_{\sigma;i}\vert\psi_{\sigma;i}\rangle,\;\, i=1,2,\dots\,.
\end{equation}
In doing so, $\mathbb{G}_{\sigma}(\varepsilon)$ denotes the matrix for which one has
\begin{equation}\label{e7}
\big(\mathbb{G}_{\sigma}(\varepsilon)\big)_{i,j} \equiv \langle\psi_{\sigma;i}\vert \h{G}_{\sigma}(\varepsilon)\vert\psi_{\sigma;j}\rangle.
\end{equation}
Thus, in principle the single-particle excitation energies of the interacting system are those $\varepsilon$ for which
\begin{equation}\label{e8}
\det\hspace{-1.4pt}\big(\mathbb{G}_{\sigma}^{-1}(\varepsilon)\big) = 0,
\end{equation}
which, in the light of the expression in Eq.\,(\ref{e5}), leads to
\begin{equation}\label{e9}
\det\hspace{-1.4pt}\big(\varepsilon\hspace{0.4pt} \mathbb{I} -\mathbb{D}_{\sigma} + \mathbb{V}_{\sigma}^{\textrm{xc}} - \hbar\hspace{0.4pt} \mathbb{S}_{\sigma}^{\textsc{g}v} - \hbar\hspace{0.4pt} \mathbb{S}_{\sigma}^{\textsc{gw}'\hspace{-2.0pt}}(\varepsilon)\big) = 0,
\end{equation}
where $\mathbb{I}$ is the unit matrix, $\mathbb{D}_{\sigma} \doteq \mathrm{diag}(\varepsilon_{\sigma;1},\varepsilon_{\sigma;2},\dots)$, and $\mathbb{V}_{\sigma}^{\textrm{xc}}$, $\mathbb{S}_{\sigma}^{\textsc{g}v}$ and $\mathbb{S}_{\sigma}^{\textsc{gw}'\hspace{-2.0pt}}(\varepsilon)$ are the Kohn-Sham matrix representations of respectively $\h{v}_{\sigma}^{\textrm{xc}}$, $\h{\Sigma}_{\sigma}^{\textsc{g}v}$ and $\h{\Sigma}_{\sigma}^{\textsc{gw}'\hspace{-2.0pt}}(\varepsilon)$.

Now enter the formalism of Neuhauser \emph{et al.} \cite{NGAKRB14a} (\emph{below Eq.\,($n$)$_{\textsc{n}}$ will refer to Eq.\,($n$) of this reference}). Solving the equation in Eq.\,(1)$_{\textsc{n}}$ in the limit of $\upsigma \to 0$, where $\upsigma^2$ is the variance of the distribution function $f_{\upsigma}(\varepsilon) \doteq \e^{-\varepsilon^2/2\upsigma^2}$  \cite{NGAKRB14a} ($\upsigma$ is not to be confused with the spin index $\sigma$), is tantamount to solving the equation in Eq.\,(\ref{e9}) above under the assumption that $\mathbb{V}_{\sigma}^{\textrm{xc}} -\hbar\hspace{0.4pt} \mathbb{S}_{\sigma}^{\textsc{g}v} -\hbar\hspace{0.4pt} \mathbb{S}_{\sigma}^{\textsc{gw}'\hspace{-2.0pt}}(\varepsilon)$ were diagonal. This is readily appreciated by expressing the trace operations in the expressions in Eq.\,(2)$_{\textsc{n}}$ as the sum over the diagonal elements of the Kohn-Sham matrix representations of the single-particle operators in these expressions; for sufficiently small $\upsigma$ and for $\varepsilon$ in a region where the Kohn-Sham eigenvalues are sufficiently apart, owing to $(f_{\upsigma}(\h{h}_{\textsc{ks}} -\varepsilon))^2$ the functions $\Sigma^{\textsc{p}}(t;\varepsilon)$, $\Sigma^{\textsc{x}}(\varepsilon)$ and $\Sigma^{\textsc{xc}}(\varepsilon)$ in Eq.\,(2)$_{\textsc{n}}$ are essentially the diagonal elements of the matrix representations of the respective operators with respect to the Kohn-Sham single-particle eigenstate whose corresponding eigenvalue is nearest to $\varepsilon$ (in the case of degeneracy, one has an average over the manifold of the relevant Kohn-Sham states). It follows that,  \emph{unless $\mathbb{V}_{\sigma}^{\textrm{xc}} -\hbar\hspace{0.4pt}\mathbb{S}_{\sigma}^{\textsc{g}v} -\hbar\hspace{0.4pt} \mathbb{S}_{\sigma}^{\textsc{gw}'\hspace{-2.0pt}}(\varepsilon)$ is diagonal, in principle the formalism of Neuhauser \emph{et al.} \cite{NGAKRB14a} does not yield the sought-after quasi-particle energies.}

For some systems (\textsl{bulk} semiconductors) it has been found that for $\varepsilon$ not far from the chemical potential, the off-diagonal elements of $\mathbb{V}_{\sigma}^{\textrm{xc}} -\hbar\hspace{0.4pt}\mathbb{S}_{\sigma}^{\textsc{g}v} -\hbar\hspace{0.4pt}\mathbb{S}_{\sigma}^{\textsc{gw}'\hspace{-2.0pt}}(\varepsilon)$ may be neglected \cite[\S\hspace{0.0pt}IV.A]{HL86}, however this is not a general principle to be relied upon, as evidenced by the contrary observations in for instance Refs.\,\cite{RL00} and \cite{PRODSB01}.

Since the eigenstates of $\h{h}_{\sigma}[\{n_{\sigma'}\}]$ are independent of $\varepsilon$, even if $\mathbb{V}_{\sigma}^{\textrm{xc}} -\hbar\hspace{0.4pt}\mathbb{S}_{\sigma}^{\textsc{g}v} -\hbar\hspace{0.4pt} \mathbb{S}_{\sigma}^{\textsc{gw}'\hspace{-2.0pt}}(\varepsilon)$ is diagonal at one specific value of $\varepsilon$, it cannot be so in a neighbourhood of this $\varepsilon$. More generally, for the cases where $\mathbb{V}_{\sigma}^{\textrm{xc}} -\hbar\hspace{0.4pt}\mathbb{S}_{\sigma}^{\textsc{g}v} -\hbar\hspace{0.4pt} \mathbb{S}_{\sigma}^{\textsc{gw}'\hspace{-2.0pt}}(\varepsilon)$ is approximately diagonal for $\varepsilon = \varepsilon_0$, the range of $\varepsilon -\varepsilon_0$ over which this approximate diagonality is maintained depends on the magnitude of the off-diagonal elements of $\partial\hbar\hspace{0.4pt} \mathbb{S}_{\sigma}^{\textsc{gw}'\hspace{-2.0pt}}(\varepsilon)/\partial\varepsilon$ at $\varepsilon = \varepsilon_0$, assuming that this derivative is bounded. In this connection, we refer the reader to Fig.\,1 of Ref.\,\cite{NGAKRB14a}, where the data displayed in the two panels clearly suggest that this is in general not the case. We remark that the very `jittery' behaviour of the functions displayed in this figure is essentially, if not entirely, due to the bounded nature of the system to which they correspond. This is relevant, in that it shows that \emph{for \textsl{finite} systems, the diagonal approximation of the equation in Eq.\,(\ref{e8}) is almost never justified} (the off-diagonal elements of $\mathbb{S}_{\sigma}^{\textsc{gw}'\hspace{-2.0pt}}(\varepsilon)$ vary similarly as their diagonal counterparts for variations of $\varepsilon$). For \textsl{extended} systems (see the data in, e.g., Figs.\, 16 and 17 of Ref.\,\cite[pp.\,86, 87]{HL69}, as well as those in Figs.\,8 and 9 of Ref.\,\cite{EFNM91}, which also reflect the consequences of both the dimensionality of the space, $d$, and the bandstructure), from the above observations it follows that for $\varepsilon$ in a neighbourhood of in particular the chemical potential, \emph{this diagonal approximation is at best valid for weakly-correlated GSs} (not for heavy-fermion systems \cite{PF95}, for instance).

One may consider the general non-negligibility of the off-diagonal elements of $\mathbb{V}_{\sigma}^{\textrm{xc}} -\hbar\hspace{0.4pt}\mathbb{S}_{\sigma}^{\textsc{g}v} -\hbar\hspace{0.4pt} \mathbb{S}_{\sigma}^{\textsc{gw}'\hspace{-2.0pt}}(\varepsilon)$ from the following alternative, albeit limited, perspective. For $\vert\varepsilon\vert\to \infty$, to leading order $\mathbb{S}_{\sigma}^{\textsc{gw}'\hspace{-2.0pt}}(\varepsilon)$ decays towards the zero matrix like $1/\varepsilon$ \cite{BF02,BF07}, so that for sufficiently large values of $\vert\varepsilon\vert$ the equation in Eq.~(\ref{e9}) can be expressed as the following asymptotic equation:
\begin{equation}\label{e10}
\det\hspace{-1.4pt}\big(\varepsilon\hspace{0.4pt} \mathbb{I} -\mathbb{D}_{\sigma} + \mathbb{V}_{\sigma}^{\textrm{xc}} - \hbar\hspace{0.4pt} \mathbb{S}_{\sigma}^{\textsc{g}v}\big) \sim 0.
\end{equation}
With $\langle\bm{r}\vert \h{v}_{\sigma}^{\textrm{xc}}[\{n_{\sigma'}\}]\vert\bm{r}'\rangle = v_{\sigma}^{\textrm{xc}}(\bm{r};[\{n_{\sigma'}\}])\hspace{0.6pt} \delta(\bm{r}-\bm{r}')$, one has
\begin{equation}\label{e11}
\big(\mathbb{V}_{\sigma}^{\textrm{xc}}\big)_{i,j} = \int \mathrm{d}^dr\;  v_{\sigma}^{\textrm{xc}}(\bm{r};[\{n_{\sigma'}\}])\hspace{0.6pt} \Psi_{\sigma;i,j}(\bm{r}),
\end{equation}
where
\begin{equation}\label{e12}
\Psi_{\sigma;i,j}(\bm{r}) \doteq \psi_{\sigma;i}^*(\bm{r}) \psi_{\sigma;j}(\bm{r}).
\end{equation}
With $f_{\sigma;i}$ denoting the occupation number of the Kohn-Sham eigenstate $\vert\psi_{\sigma;i}\rangle$ in the $N$-particle Kohn-Sham GS of the system under investigation, for the Kohn-Sham density-matrix operator $\h{\varrho}_{\sigma}$ one has $\h{\varrho}_{\sigma} = \sum_{k} f_{\sigma;k} \vert\psi_{\sigma;k}\rangle\langle\psi_{\sigma;k}\vert$, from which and from the expression in Eq.\,(\ref{e2}) one obtains
\begin{eqnarray}\label{e13}
\big(\mathbb{S}_{\sigma}^{\textsc{g}v}\big)_{i,j} &\equiv& \langle\psi_{\sigma;i}\vert \h{\Sigma}_{\sigma}^{\textsc{g}v}\vert\psi_{\sigma;j}\rangle
= -\frac{1}{\hbar}\sum_k f_{\sigma;k}\nonumber\\
&\times& \int \textrm{d}^dr \textrm{d}^dr'\; \Psi_{\sigma;i,k}(\bm{r})\hspace{0.4pt} v(\bm{r}-\bm{r}')\hspace{0.4pt} \Psi_{\sigma;j,k}^*(\bm{r}'). \hspace{0.6cm}
\end{eqnarray}
We note in passing that within the framework of the local-density approximation \cite{KS65,vBH72}, the exchange-correlation potential $v_{\sigma}^{\textrm{xc}}(\bm{r};[\{n_{\sigma'}\}])$ is negative for all $\bm{r}$, implying that at least in this approximation the diagonal elements of $\mathbb{V}_{\sigma}^{\textrm{xc}}$ are negative. The two-body Coulomb potential $v(\bm{r}-\bm{r}')$ is evidently positive. For $i = j = k$, the integral on the right-hand side of Eq.\,(\ref{e13}) is a specific form of the so-called \textsl{Coulomb integral}, $C_{ii}^{\sigma}$, which is clearly positive, and for $i=j \not= k$, it is the so-called \textsl{exchange integral}, $J_{ik}^{\sigma}$, which can be shown to be also positive \cite[p.\,30]{PF03}. The constant $C_{ii}^{\sigma}$ is what one would call \textsl{on-site Coulomb interaction}, $U_{ii}^{\sigma}$, for the cases where $\psi_{\sigma;i}(\bm{r})$ is centred on an atomic position indexed $i$ \cite[p.\,69]{PF03}.

Heuristically, from the above expressions one observes that in order for $\mathbb{V}_{\sigma}^{\textrm{xc}}$ to be diagonal, one should have $\Psi_{\sigma;i,j}(\bm{r}) \propto \delta_{i,j}$, and in order for $\mathbb{S}_{\sigma}^{\textsc{g}v}$ to be diagonal, $\Psi_{\sigma;i,k}(\bm{r}) \propto \delta_{i,k}$ for all $k$ for which $f_{\sigma;k}\not=0$. The former property encompasses the latter one. The property $\Psi_{\sigma;i,j}(\bm{r}) \propto \delta_{i,j}$, for all $i,j$, is clearly \textsl{not} realized in general, however it may in principle be realized in a model system where $\psi_{\sigma;i}(\bm{r})$ is strongly localized around the lattice point $\bm{R}_i$, for all $i$. One may think of $\{\psi_{\sigma;i}(\bm{r}) \| i\}$ as ideally corresponding to the half-filled GS of a lattice of atoms, located at $\{\bm{R}_i \| i\}$, in the limit of infinite on-site repulsion energy of electrons (or zero hopping amplitude, corresponding to the atomic limit) \cite[Ch.\,5]{PF03}. In this strong-coupling limit, the energy of the $N$-particle GS of $\wh{H}_{\textsc{ks}}$ is highly degenerate, undermining the possibility of employing the non-self-consistent zero-temperature perturbation expansion of the self-energy operator around such GS. In this limit, not the electron-electron interaction operator, but the kinetic-energy operator, or the `band Hamiltonian', that is to be treated as perturbation, leading to strong-coupling perturbation schemes \cite[Chs\,4, 5]{PF03}. \emph{We conclude that in the region of large values of $\vert\varepsilon\vert$, the diagonal approximation of the equation in Eq.\,(\ref{e9}) is in general invalid.} Evidently, representing $\h{G}_{\sigma}^{-1}(\varepsilon)$ in terms of the eigenstates of $\h{h}_{\sigma} - \h{v}_{\sigma}^{\textrm{xc}} +  \hbar\hspace{0.4pt}\h{\Sigma}_{\sigma}^{\textsc{g}v}$, Eq.\,(\ref{e5}), the diagonal approximation of the equation in Eq.\,(\ref{e8}) becomes asymptotically exact for $\vert\varepsilon\vert\to \infty$.

We shall return to the solution of the equation in Eq.\,(\ref{e9}) later in this Comment, after having discussed \emph{the second fundamental shortcoming} of the formalism of Neuhauser \emph{et al.} \cite{NGAKRB14a}. Here use has been made of the following equality \cite[Eq.\,(10)]{NGAKRB14a}:
\begin{equation}\label{e14}
\langle\psi\vert \h{v} \otimes \h{\chi}^{\textrm{r}}(t) \otimes \h{v} \vert\zeta\phi\rangle = \langle\psi\vert\h{v}\vert\delta n(t)\rangle,
\end{equation}
where $\h{v}$ and $\h{\chi}^{\textrm{r}}(t) = \sum_{\sigma} \h{\chi}_{\sigma}^{\textrm{r}}(t)$ denote the single-particle operators for respectively the two-body Coulomb potential (denoted by $u_{\textsc{c}}$ in Ref.\,\cite{NGAKRB14a}) and the retarded \textsl{interacting} density-density response function \cite[Ch.\,5, \S13]{FW03} ($\h{\chi}^{\textrm{r}}(t)$ is to be distinguished from its Kohn-Sham counterpart, $\h{\chi}_{\textsc{ks}}^{\textrm{r}}(t)$), and $\otimes$ signifies ``space convolution'' \cite{NGAKRB14a}. Neuhauser \emph{et al.} \cite{NGAKRB14a,NGAKRB14b} evaluate the vector $\vert\delta n(t)\rangle$ by integrating the time-dependent single-particle Kohn-Sham equation \cite[Ch.\,7]{ED11} subject to the perturbation $\delta v_{\textrm{ext}}(\bm{r},t) \doteq \langle\bm{r}\vert \h{v}\vert\zeta\phi\rangle\hspace{0.4pt} \delta(t)$. The kernel of the time-dependent Kohn-Sham equation integrated in the relevant calculations \cite{NGAKRB14b}, consists of the Kohn-Sham Hamiltonian $\h{h}_{\sigma}[\{n_{\sigma'}\}]$ corresponding to the GS of the system under investigation, Eq.\,(\ref{e3}), supplemented with the Hartree potential associated with the time-dependent total number-density fluctuation $\delta n(\bm{r},t) \equiv \sum_{\sigma} \delta n_{\sigma}(\bm{r},t)$ \cite[Eq.\,(5)]{NGAKRB14b}. Now, whereas the equality in Eq.\,(\ref{e14}) is correct (by definition), the adopted method of calculating $\vert\delta n(t)\rangle$ \textsl{is not}. This is because the function $\langle\bm{r}\vert\delta n(t)\rangle$ as calculated by Neuhauser \emph{et al.} \cite{NGAKRB14a,NGAKRB14b} takes account of $\langle\bm{r}\vert\h{v}\vert\zeta\phi\rangle$ (the amplitude of the perturbing $\delta$-function pulse at $t=0$) to \textsl{all} orders, and not to \textsl{linear} order. Although admittedly the calculated $\langle\bm{r}\vert\delta n(t)\rangle$ describes the temporal evolution of the \textsl{physical} total number density of the interacting system, in response to the perturbation applied at time $t=0$, this response is \textsl{not} the one to be taken account of in the calculation of the $GW$ self-energy operator: the operator $\h{\chi}(t)$, the retarded part of which, $\h{\chi}^{\textrm{r}}(t)$, one encounters in Eq.\,(\ref{e14}) above, is the coefficient of the \textsl{linear} term in the functional expansion of the time-dependent total number density in powers of the time-dependent variation in the \textsl{external} potential; as such, it must be fully independent of the latter perturbing potential. In this connection, it is important to realize that since in the case at hand the perturbation, that is $\langle\bm{r}\vert\h{v} \vert\zeta\phi\rangle\hspace{0.4pt} \delta(t)$, is \textsl{not} weak, the non-linear effects taken into account in the formalism of Neuhauser \emph{et al.} \cite{NGAKRB14a} \textsl{cannot} be negligible. Consequently, the calculation of the number-density response function as appropriate in the context of the determination of $\h{\Sigma}_{\sigma}^{\textsc{gw}'\hspace{-2.0pt}}(\varepsilon)$ is to be performed along the lines of the \textsl{linear-response} formalism by Baer and Neuhauser \cite{BN04}, discussed further by Neuhauser and Baer in Ref.\,\cite{NB05}. Clearly, the repeated matrix-vector multiplications to be carried out in this formalism (involving the $2\hspace{0.5pt}\mathcal{N} \times 2\hspace{0.5pt}\mathcal{N}$ matrix $\mathbb{A}$ \cite[Eq.\,(17)]{BN04}, where $2\hspace{0.5pt}\mathcal{N}$ is the total number of electrons) render the arithmetic complexity of the formalism scaling at least like $\mathcal{N}^2$, leaving aside the instabilities that are inherent in such calculations \cite{BN04}.

We note in passing that use of the above-mentioned Hartree approximation is consistent with the evaluation of the zero-order polarization diagram \cite{JH57,FW03} (coinciding with the Kohn-Sham density-density response operator $\h{\chi}_{\textsc{ks}}$, referred to above) in the calculation of the dielectric response function within the framework of the random-phase approximation, RPA. The incorporation of for instance the adiabatic approximation of the exchange-correlation kernel \cite[\S\S7.3, 7.4]{ED11} would change this picture. It would however result in the $\h{\chi}$, calculated from
\begin{equation}
\h{\chi} = \h{\chi}_{\textsc{ks}} + \h{\chi}_{\textsc{ks}} (\h{v} + \h{f}_{\textrm{xc}}) \h{\chi} \Leftrightarrow \h{\chi} = (\h{1} - \h{\chi}_{\textsc{ks}} [\h{v}+\h{f}_{\textrm{xc}}])^{-1} \h{\chi}_{\textsc{ks}},
\nonumber
\end{equation}
coinciding with the \textsl{exact} $\h{\chi}$ in the static limit, that is in the limit of $\varepsilon\to 0$ \cite[\S8.6.1, p.\,194]{BF99a}, \cite{BF99b} (see however the discussions in Ref.\,\cite[\S6.7, p.\,167]{BF99a}). We further note that calculation of the time-dependent number density, as required for the evaluation of the self-energy, can be relatively straightforwardly accomplished through solving the (linearized) Liouville equation of motion for the single-particle density operator \cite[Ch.\,6]{LER98}, along the lines described by Ehrenreich and Cohen \cite{EC59}.

\vspace{0.0cm}
\emph{Some relevant technical details.} For the discussions to be presented below, we consider the coordinate representations of dynamic operators in the complex energy ($z$) plane; the choice of coordinate representation is in part motivated by the approach in Refs.\,\cite{NGAKRB14a,NGAKRB14b}. To distinguish these functions from their `physical' counterparts, on the real axis of the $z$ plane (that is, the $\varepsilon$ axis), we furnish the functions of $z$ with a tilde. Thus, with $\Sigma_{\sigma}^{\textsc{gw}}(\bm{r},\bm{r}';\varepsilon)$ denoting the coordinate representation of $\h{\Sigma}_{\sigma}^{\textsc{gw}}(\varepsilon)$, $\varepsilon \in \mathds{R}$,  $\t{\Sigma}_{\sigma}^{\textsc{gw}}(\bm{r},\bm{r}';z)$ denotes the analytic continuation of $\Sigma_{\sigma}^{\textsc{gw}}(\bm{r},\bm{r}';\varepsilon)$ into the complex $z$ plane. The two functions are related as follows:
\begin{equation}\label{e15}
\Sigma_{\sigma}^{\textsc{gw}}(\bm{r},\bm{r}';\varepsilon) = \lim_{\eta\downarrow 0} \t{\Sigma}_{\sigma}^{\textsc{gw}}(\bm{r},\bm{r}';\varepsilon \pm i\eta),\;\, \varepsilon \gtrless \mu,
\end{equation}
where $\mu$ is the chemical potential. In a non-self-consistent calculation, such as relevant to the considerations of Ref.\,\cite{NGAKRB14a}, this chemical potential coincides with that corresponding to the underlying Kohn-Sham GS. This should however \textsl{not} suggest that this $\mu$ were to be identified as the thermodynamic chemical potential within the framework of the $GW$ approximation. In this connection, we note that a non-self-consistently calculated $\t{\Sigma}_{\sigma}^{\textsc{gw}}(\bm{r},\bm{r}';z)$ and the associated single-particle Green function (the two being related through the Dyson equation) fail to satisfy the Luttinger-Ward identity \cite{BF13}. \emph{Similar expressions as that in Eq.\,(\ref{e15}) apply to $\t{G}_{\sigma}(\bm{r},\bm{r}';z)$ and $\t{W}(\bm{r},\bm{r}';z)$, except that for $\t{W}$ the role of $\mu$ is taken over by $0$ \cite{BF02,BF07}.}

The specific way in which $\Sigma_{\sigma}^{\textsc{gw}}(\bm{r},\bm{r}';\varepsilon)$ is related to $\t{\Sigma}_{\sigma}^{\textsc{gw}}(\bm{r},\bm{r}';z)$ (and similarly $G_{\sigma}(\bm{r},\bm{r}';\varepsilon)$ to $\t{G}_{\sigma}(\bm{r},\bm{r}';z)$, and $W(\bm{r},\bm{r}';\varepsilon)$ to $\t{W}(\bm{r},\bm{r}';z)$), Eq.\,(\ref{e15}), is relevant for two distinct reasons, of which one is the fact that $\t{\Sigma}_{\sigma}^{\textsc{gw}}(\bm{r},\bm{r}';z)$ in general undergoes branch-cut discontinuity along at least some continuous parts of the real $\varepsilon$ axis on $z$ crossing this axis, rendering in general the function $\t{\Sigma}_{\sigma}^{\textsc{gw}}(\bm{r},\bm{r}';\varepsilon)$, with $\varepsilon\in\mathds{R}$, ambiguous. The second reason, which is perhaps less highlighted in the literature, is related to the specific way in which $z$ is to approach the $\varepsilon$ axis, namely from the lower/upper-half part of the $z$ plane for $\re(z) \lessgtr \mu$.

To clarify the latter aspect, we consider the Green function $G_0$ as represented in Eq.\,(4) of Ref.\,\cite{NGAKRB14a} in the limit of $\beta\to\infty$ (see the definition of the distribution $\h{P}_{\mu}(t)$ in Ref.\,\cite{NGAKRB14a}). Making use of the closure relation $\sum_i \vert\psi_{\sigma;i}\rangle \langle\psi_{\sigma;i}\vert = \h{1}$ for the eigenstates of $\h{h}_{\sigma}[\{n_{\sigma'}\}]$ (that is $\h{h}_{\textsc{ks}}$ in the notation of Ref.\,\cite{NGAKRB14a}), Eq.\,(\ref{e6}), one immediately observes that aside from the unit-step functions $\theta(\pm t)$, the dependence of $G_0(\bm{r},\bm{r}';t)$ on $t$ is determined by the \textsl{undamped} oscillatory function $\e^{-i \varepsilon_{\sigma;i} t/\hbar}$. This implies that the Fourier transformation of this function with respect to $t$ \textsl{cannot} be effected directly. Instead, as we have emphasized in Ref.\,\cite[\S4, p.\,125]{BF99a}, the integral $\int_{-\infty}^{\infty} \rd t\, \e^{i\varepsilon t/\hbar} (\dots)$ must first be decomposed as $\int_{-\infty}^{0} \rd t\, \e^{i\varepsilon t/\hbar} (\dots) + \int_{0}^{\infty} \rd t\, \e^{i\varepsilon t/\hbar} (\dots)$, giving rise to the `hole' and `particle' Green functions, $G_{\sigma}^{\textrm{h}}$ and $G_{\sigma}^{\textrm{p}}$ respectively \cite[Eq.\,(10)]{BF99a}. These functions, expressed as integrals with respect to $t$, initially exist only for $\varepsilon = z \in \mathds{C}$, with $\im(z)\lessgtr 0$ in the case of $G_{\sigma}^{\textrm{h}/\textrm{p}}$. Once the relevant integrals with respect to $t$ have been evaluated, the resulting functions $\t{G}_{\sigma}^{\textsl{h}/\textrm{p}}(\bm{r},\bm{r}';z)$ can be analytically continued to the half planes $\im(z)\gtrless 0$ (cf. \cite[\S6.3.4, p.\,105]{BF07}). Thus one obtains the function $\t{G}_{\sigma}(\bm{r},\bm{r}';z) \equiv \t{G}_{\sigma}^{\textsl{h}}(\bm{r},\bm{r}';z) + \t{G}_{\sigma}^{\textsl{p}}(\bm{r},\bm{r}';z)$, which is defined everywhere on the complex $z$ plane where it is bounded, specifically in the region $\im(z) \not=0$. Since on general grounds $\t{G}_{\sigma}^{\textsl{h}/\textrm{p}}(\bm{r},\bm{r}';z)$ are analytic in the regions $\re(z)\gtrless \mu$ (a fact that can be explicitly verified), from the above observations one arrives at the above specification for obtaining $G_{\sigma}(\bm{r},\bm{r}';\varepsilon)$, $\varepsilon \in\mathds{R}$, from $\t{G}_{\sigma}(\bm{r},\bm{r}';z)$, $\im(z) \not=0$ (cf. Eq.\,(\ref{e15})).

Having presented the above auxiliary details, we are now in a position to state that on account of the time-reversal symmetry of the GS \cite[\S4.4, p.\,132]{BF99a}, one has
\begin{equation}\label{e16}
\t{\Sigma}_{\sigma}^{\textsc{gw}'\hspace{-2.0pt}}(\bm{r},\bm{r}';z^*) = \t{\Sigma}_{\sigma}^{\textsc{gw}'\hspace{-1.0pt}*}(\bm{r},\bm{r}';z),\;\; \im(z) \not= 0.
\end{equation}
Further \cite[cf. Eqs.\,(B.55), (B.59)]{BF07}
\begin{equation}\label{e17}
\sgn(\im[\t{\Sigma}_{\sigma}^{\textsc{gw}'\hspace{-2.0pt}}(\bm{r},\bm{r}';z)]) = -\sgn(\im(z)),\;\, \im(z)\not=0.
\end{equation}
This result is significant in that it shows that with the exception of some possible isolated points on the real $\varepsilon$ axis, the equation in Eq.\,(\ref{e9}), as well as its approximation arrived at through neglecting the off-diagonal elements of $\mathbb{V}_{\sigma}^{\textrm{xc}} -\hbar\hspace{0.4pt}\mathbb{S}_{\sigma}^{\textsc{g}v}- \hbar\hspace{0.4pt} \mathbb{S}_{\sigma}^{\textsc{gw}'\hspace{-2.0pt}}(\varepsilon)$, Eq.\,(\ref{e18}), has \textsl{no} solution on the physical Riemann sheet \cite[\S6, p.\,145]{BF99a} (see later). This fact raises the question as to the mechanism whereby Neuhauser \emph{et al.} \cite{NGAKRB14a,NGAKRB14b} have solved Eq.\,(1)$_{\textsc{n}}$ (see Fig.\,2 in Ref.\,\cite{NGAKRB14a}, and note that this reference contains no mention regarding the imaginary parts of the calculated $\varepsilon$-dependent functions); as we have indicated earlier, of course we realize that the calculations presented in Ref.\,\cite{NGAKRB14a} concern \textsl{bounded} systems. Part of the answer to this question can be found under items $7$ and $11$ of Ref.\,\cite[pp.\,5, 6]{NGAKRB14b}: multiplication of functions of $t$ with the regularization function $\e^{-(\Gamma t)^2/2}$, where $t \equiv \tau \delta t$, prior to Fourier transformation. Use of this regularization function, in which $(\Gamma t)^2/2$ cannot be brought into an analytic connection with $i (\varepsilon \pm i \eta) t$, $\eta >0$, does \textsl{not} conform with the different conditions, described above, under which  $\t{G}_{\sigma}^{\textrm{p}}(\bm{r},\bm{r}';z)$ and $\t{G}_{\sigma}^{\textrm{h}}(\bm{r},\bm{r}';z)$ are calculated. Other part of the answer can be found in Sec.\,6.3 of Ref.\,\cite[p.\,97]{BF07}: use of insufficiently large cut-off energies in the numerical calculations of the self-energy leads to a noticeable degree of violation of causality. Consequences of the use of a finite $\beta \equiv 1/k_{\textsc{b}} T$, inappropriately mixing levels below and above $\mu$, is also to be reckoned with \cite[Appendix C]{BF07}.

We remark that in order to time Fourier transform products of such functions as $g(t)$ and $w(t)$, where (cf. \cite[Eq.\,(7.47)]{FW03} and \cite[Eq.\,(C.3), p.\,180]{HL69})
\begin{eqnarray}
g(t) &=& g_-(t)\hspace{0.4pt} \theta(-t) + g_+(t)\hspace{0.4pt} \theta(t),
\nonumber\\
w(t) &=& w_-(t)\hspace{0.4pt} \theta(-t) + w_+(t)\hspace{0.4pt} \theta(t),
\nonumber
\end{eqnarray}
one should first make use of the identity (neglecting the unimportant set $\{0\}$ on the $t$ axis)
\begin{equation}
g(t) w(t) \equiv g_-(t) w_-(t)\hspace{0.4pt} \theta(-t) + g_+(t) w_+(t)\hspace{0.4pt} \theta(t),
\nonumber
\end{equation}
and subsequently \textsl{separately} time Fourier transform the functions $g_-(t) w_-(t)\hspace{0.4pt} \theta(-t)$ and  $g_+(t) w_+(t)\hspace{0.4pt} \theta(t)$, along the lines described above in dealing with the functions $\t{G}_{\sigma}^{\textrm{h}}$ and $\t{G}_{\sigma}^{\textrm{p}}$. In doing so, choice of $z$ in the appropriate half of the complex $z$ plane renders use of such inappropriate regularization function as $\e^{-(\Gamma t)^2/2}$ redundant.

Considering for simplicity the diagonal approximation of the equation in Eq.\,(\ref{e9}),
\begin{eqnarray}\label{e18}
&&\hspace{-0.4cm}\varepsilon \approx \varepsilon_{\sigma;i} - \langle\psi_{\sigma;i}\vert \h{v}_{\sigma}^{\textrm{xc}}\vert\psi_{\sigma;i}\rangle + \hbar \langle\psi_{\sigma;i}\vert\h{\Sigma}_{\sigma}^{\textsc{g}v}\vert\psi_{\sigma;i}\rangle\nonumber\\
 &&\hspace{3.4cm} + \hbar \langle\psi_{\sigma;i}\vert \h{\Sigma}_{\sigma}^{\textsc{gw}'\hspace{-2.0pt}}(\varepsilon) \vert\psi_{\sigma;i}\rangle,
\end{eqnarray}
for some $i$, this equation can be solved (that is, solved on the non-physical Riemann sheet neighbouring the physical one \cite[\S2.2, p.\,114]{BF99a}) through employing a finite-order Taylor expansion of  $\langle\psi_{\sigma;i}\vert\h{\Sigma}_{\sigma}^{\textsc{gw}'\hspace{-2.0pt}}(\varepsilon)\vert\psi_{\sigma;i}\rangle$ around a point on the $\varepsilon$ axis, say around $\varepsilon =  \varepsilon_{\sigma;i}$ (see Ref.\,\cite{FEDvH94} and the references herein). The well-known `quasi-particle approximation' \cite[\S6.1, p.\,150]{BF99a} amounts to expanding the latter function to linear order in $(\varepsilon-\varepsilon_{\sigma;i})$. The relevant expressions are presented in, for instance, Ref.\,\cite[Eqs.\,(36), (37) ]{HL86}, and  Ref.\,\cite[Eqs.\,(34), (35)]{FS06}.

\vspace{0.0cm}
In conclusion, we have discussed that an approximation inherent in the formalism of Neuhauser \emph{et al.} \cite{NGAKRB14a}, namely the diagonal approximation of the equation for quasi-particle energies, is in general invalid. Further, we have shown that the purported self-energy function as calculated in Ref.\,\cite{NGAKRB14a}, fundamentally deviates from the intended $GW$ approximation of the self-energy operator, $\h{\Sigma}_{\sigma}^{\textsc{gw}}(\varepsilon)$. On employing the appropriate equation of motion specific to calculating the interacting \textsl{linear} density-density response function, required for the calculation of $\h{\Sigma}_{\sigma}^{\textsc{gw}}(\varepsilon)$, the arithmetic complexity of the stochastic formalism put forward by Neuhauser \emph{et al.} \cite{NGAKRB14a} will at best scale \textsl{quadratically} with the number of electrons in the system. \hfill$\square$

\vfill
%

\bibliographystyle{apsrev}

\begin{thebibliography}{10}

\bibitem{NGAKRB14a}
D. Neuhauser, Y. Gao, C. Arntsen, C. Karshenas, E. Rabani, and R. Baer, \emph{Breaking the theoretical scaling limit for predicting quasi-particle energies: The stochastic $GW$ approach}, \href{http://arxiv.org/abs/1402.5035}{arXiv:1402.5035v1}.

\bibitem{NGAKRB14b}
D. Neuhauser, Y. Gao, C. Arntsen, C. Karshenas, E. Rabani, and R. Baer, \href{http://www.fh.huji.ac.il/~roib/Postscripts/sGW-SI.pdf}{\emph{Supporting information}} for Ref.\,\protect\cite{NGAKRB14a}.

\bibitem{LH65}
L. Hedin, \emph{Phys. Rev.} \textbf{139}, A\,796 (1965).

\bibitem{HL69}
L. Hedin, and S. Lundqvist, in \emph{Solid State Physics}, Vol.\,\textbf{23}, edited by F. Seitz, D. Turnbull and H. Ehrenreich (Academic Press, New York, 1969).

\bibitem{DL94}
S.-J. Dong, and K.-F. Liu, \emph{Phys. Lett.} B\,\textbf{328}, 130 (1994).

\bibitem{KP05}
F.\,R. Krajewski, and M. Parrinello, \emph{Phys. Rev.} B\,\textbf{71}, 233105 (2005).

\bibitem{BF14}
B. Farid, \textsl{An unconstrained order-$\mathcal{N}$ scaling density-functional formalism -- with some digressions concerning efficient many-body calculations}, to be published. \\ Briefly, in order for the relative error in the calculated Kohn-Sham kinetic energy to be kept under control (when using the stochastic method of matrix inversion), the number of noise vectors to be employed must scale like $\mathcal{N}^{\alpha}$ with $\alpha > 2$, where $\mathcal{N}$ is the number of atoms, or equivalently electrons, in the system under consideration. The strict condition $\alpha >2$ renders the stochastic method more time-consuming than the standard order-$\mathcal{N}^3$ method of electronic-structure calculations. Formally, the two methods would be of comparable arithmetic complexity for $\alpha = 2$, although in this case the pre-factor of $\mathcal{N}^3$ would still be larger for the stochastic method.

\bibitem{JH57}
J. Hubbard, \emph{Proc. Roy. Soc. London}, A \textbf{240}, 539 (1957).

\bibitem{KS65}
W. Kohn, L.\,J. Sham, \emph{Phys. Rev.} \textbf{140}, A\,1133 (1965).

\bibitem{vBH72}
U. von Barth, L. Hedin, \emph{J. Phys.} C\,\textbf{5}, 1629 (1972).

\bibitem{FW03}
A.\,L. Fetter, J.\,D. Walecka, \emph{Quantum Theory of Many-Particle Systems} (Dover, New York, 2003).

\bibitem{HL86}
M.\,S. Hybertsen, and S.\,G. Louie, \emph{Phys. Rev.} \textbf{34}, 5390 (1986).

\bibitem{RL00}
M. Rohlfing, and S.\,G. Louie, \emph{Phys. Rev.} B\,\textbf{62}, 4927 (2000).

\bibitem{PRODSB01}
O. Pulci, L. Reining, G. Onida, R. Del Sole, and F. Bechstedt, \emph{Com. Mat. Sci.} \textbf{20}, 300 (2001).

\bibitem{EFNM91}
G.\,E. Engel, B. Farid, C.\,M.\,M. Nex, and N.\,H. March, \emph{Phys. Rev.} B\,\textbf{44}, 13356 (1991).


\bibitem{PF95}
P. Fulde, \emph{Electron Correlations in Molecules and Solids}, 3rd edition (Springer, Berlin, 1995).

\bibitem{BF02}
B. Farid, \emph{Phil. Mag.} B\,\textbf{82}, 1413 (2002).

\bibitem{BF07}
B. Farid, \textsl{On the Luttinger theorem concerning number of particles in the ground states of systems of interacting fermions}, \href{http://arxiv.org/abs/0711.0952}{arXiv:0711.0952v1}.

\bibitem{PF03}
P. Fazekas, \emph{Lecture Notes on Electron Correlation and Magnetism}, Series in Modern Condensed Matter Physics, Vol.\,\textbf{5} (World Scientific, Singapore, 2003).

\bibitem{ED11}
E. Engel, and R.\,M. Dreizler, \emph{Density Functional Theory: An Advanced Course} (Springer, Berlin, 2011).

\bibitem{BN04}
R. Baer, and D. Neuhauser, \emph{J. Chem. Phys.} \textbf{121}, 9803 (2004).

\bibitem{NB05}
D. Neuhauser, and R. Baer, \emph{J. Chem. Phys.} \textbf{123}, 204105 (2005).

\bibitem{BF99a}
B. Farid, in \emph{Electron Correlation in the Solid State}, edited by N.\,H. March (Imperial College Press, London, 1999). Ch.\,3, pp.\,103-261.

\bibitem{BF99b}
B. Farid, \emph{Phys. Mag. Lett.} \textbf{79}, 581 (1999).

\bibitem{LER98}
L.\,E. Reichl, \emph{A Modern Course in Statistical Physics}, 2nd edition (John Wiley \& Sons, New York, 1998).

\bibitem{EC59}
H. Ehrenreich, and M.\,H. Cohen, \emph{Phys. Rev.} \textbf{115}, 786 (1959).

\bibitem{BF13}
B. Farid, \emph{Some rigorous results concerning the uniform ground states of single-band Hamiltonians in arbitrary dimensions}, \href{http://arxiv.org/abs/1305.2089}{arXiv:1305.2089v1}.

\bibitem{FEDvH94}
B. Farid, G.\,E. Engel, R. Daling, and W. van Haeringen, \emph{Phil. Mag.} B\,\textbf{69}, 901 (1994).

\bibitem{FS06}
C. Friedrich, and A. Schindlmayr, NIC Series, Vol.\,\textbf{31}, edited by J. Grotendorst, S. Bl\"{u}gel, and D. Marx, 335 (2006).

\end{thebibliography}
\end{document}